\documentclass[preprints,article,accept,moreauthors,pdftex]{Definitions/mdpi}

\firstpage{1}
\pubvolume{xx}
\issuenum{x}
\articlenumber{x}
\pubyear{2022}
\copyrightyear{2022}
\history{Received: date; Accepted: date; Published: date}

\usepackage{color,xspace}
\usepackage{Definitions/macros_gpi}
\usepackage{amssymb}
\usepackage{hyperref}
\usepackage{cleveref}
\usepackage{tikz}
\usepackage{pdflscape}

\graphicspath{{Figs/}}

\setitemize{parsep=6pt,itemsep=0pt,leftmargin=*,labelsep=5.5mm}
\setenumerate{parsep=6pt,itemsep=0pt,leftmargin=*,labelsep=5.5mm} 
\setlist[description]{itemsep=0mm}   
\usepackage{environ}
\NewEnviron{myequation}{%
\begin{equation} 
\scalebox{0.95}{$\BODY$}
\end{equation}
}

\Title{ Marginal Bayesian Statistics Using Masked Autoregressive Flows and Kernel Density Estimators with Examples in Cosmology$^\dagger$}

\Author{Harry Bevins $^{1, 2}$, Will Handley $^{1, 2}$,
Pablo Lemos $^{3, 4}$, Peter Sims $^{5}$
,
Eloy de Lera Acedo $^{1, 2}$, Anastasia Fialkov $^{2, 3}$} 

\AuthorNames{Harry T. J. Bevins, William J. Handley, Pablo Lemos, Peter Sims, Eloy de Lera Acedo, Anastasia Fialkov}

\address{%
$^1$ \quad Astrophysics Group, Cavendish Laboratory, J. J. Thomson Avenue, Cambridge, CB3 0HE, UK \\
$^{2}$ \quad Kavli Institute for Cosmology, Madingley Road, Cambridge CB3 0HA, UK \\
$^{3}$ \quad Department of Physics \& Astronomy, University College London, Gower Street, London, WC1E 6BT, UK \\
$^{4}$ \quad Department of Physics and Astronomy, Pevensey Building, University of Sussex, Brighton, BN1 9QH, UK \\
$^{5}$ \quad Department of Physics and McGill Space Institute, McGill University, 3600 University Street, Montreal, QC H3A 2T8, Canada \\
$^{6}$ \quad Institute of Astronomy, University of Cambridge, Madingley Road, Cambridge CB3 0HA, UK \\
}

\firstnote{Submitted to International Workshop on Bayesian Inference and Maximum Entropy Methods in Science and Engineering, IHP, Paris, July 18-22, 2022.} 

\abstract{ 
Cosmological experiments often employ Bayesian workflows to derive constraints on cosmological and astrophysical parameters from their data. It has been shown that these constraints can be combined across different probes such as Planck and the Dark Energy Survey and that this can be a valuable exercise to improve our understanding of the universe and quantify tension between multiple experiments.
However, these experiments are typically plagued by differing systematics, instrumental effects and contaminating signals, which we collectively refer to as `nuisance' components, that have to be modelled alongside target signals of interest. This leads to high dimensional parameter spaces, especially when combining data sets, with $ \gtrsim 20$ dimensions of which only $\sim 5$ correspond to key physical quantities. We present a means by which to combine constraints from different data sets in a computationally efficient manner by generating rapid, reusable and reliable marginal probability density estimators, giving us access to nuisance-free likelihoods. This is possible through the unique combination of nested sampling, which gives us access to Bayesian evidences, and the marginal Bayesian statistics code \textsc{margarine}. Our method is lossless in the signal parameters, resulting in the same posterior distributions as would be found from a full nested sampling run over all nuisance parameters, and typically quicker than evaluating full likelihoods. We demonstrate our approach by applying it to the combination of posteriors from the Dark Energy Survey and Planck.
}


\begin{document}


\section{Introduction}
\label{sec:Introduction}

Bayesian inference is a cornerstone of modern cosmology and astrophysics. It is frequently employed to derive parameter constraints on key signal parameters from data sets such as the Dark Energy Survey~\citep[DES, ][]{DES_Year1_2018, DES_year3_2021}, Planck \cite{Planck_cosmo_2020}, REACH \cite{REACH} and SARAS2 \cite{SARAS2} among others. Often experiments are sensitive to different aspects of the same physics, and by combining constraints across probes we can improve our understanding of the Universe or reveal tensions between different experiments.

However, this can become a computationally expensive task as many experiments feature systematics, instrumental effects and contamination from other physical signals that need to be model alongside the signal or parameters of interest. For individual experiments, this can lead to high dimensional problems with $\gtrsim 20$ parameters of which the majority can be considered `nuisance' parameters. The problem is compounded when combining different data sets with different models for common nuisance components and different systematics or instrumental effects that have to be modelled.

In this work we demonstrate that we can use density estimators, such as Kernel Density Estimators \cite{parzen_KDE_1962, rosenblatt_KDE_1956} and Masked Autoregressive Flows \cite{Papamarkarios_MAF_2017}, to rapidly calculate reliable and reusable representations of marginal probability densities and marginal Bayesian summary statistics for key signal or cosmological parameters. This gives us access to the nuisance-free likelihood functions and allows us to combine parameter constraints from different data sets in a computationally efficient manner given marginal posterior samples from the different experiments. We use the publicly available code\footnote{https://github.com/htjb/margarine} \textsc{margarine} \cite{margarine_neurips} to generate density estimators.

In \cref{sec:theory} we mathematically demonstrate that the application of \textsc{margarine} to the problem of combining the marginal posteriors from two data sets is equivalent to running a full nested sampling run including all `nuisance' parameters. We define in this section the nuisance-free likelihood. \Cref{sec:methods} briefly discusses the methodology behind \textsc{margarine} with reference to a previously published work \cite{margarine_neurips}.  Finally, we show the results of combining samples from DES and Planck in \cref{sec:results} and conclude in \cref{sec:conclusion}.

\section{Theory}
\label{sec:theory}
\subsection*{Notation} Given a likelihood $\mathcal{L}(\Theta)\equiv P(D|\Theta,\mathcal{M})$ representing the probability of data $D$ given some model $\mathcal{M}$ with parameters $\Theta$, Bayesian inference proceeds by defining a prior $\pi(\Theta)\equiv P(\Theta|\mathcal{M})$, and then through Bayes theorem computing a posterior distribution $\mathcal{P}(\Theta)\equiv P(\Theta|D,\mathcal{M})$ for the purposes of parameter estimation and an evidence $\mathcal{Z}\equiv P(D|\mathcal{M})$ in order to perform model comparison. In our notation we suppress model dependence, but where we wish to refer to the likelihoods derived from different datasets, we denote this with a subscript so for example $\mathcal{L}_A(\Theta)\equiv P(D_A|\Theta,\mathcal{M})$, and $\mathcal{L}_B(\Theta)\equiv P(D_B|\Theta,\mathcal{M})$.

In our setting, the parameter vector is split into two sub-vectors $\Theta=(\theta,\alpha)$, where $\theta$ are parameters of scientific interest, and $\alpha$ are \emph{nuisance} parameters, included for the purposes of data analysis. Such situations are common in astrophysics, where for example $\theta$ might be parameters governing the Universe's evolution, whilst $\alpha$ might be associated with instrument calibration and foreground removal \cite{Planck_cosmo_2020, REACH}. The $\alpha$ parameters are generally ``marginalised out'' and not considered further in final or future analyses.

\subsection*{Definitions}
With this notation established, the version of Bayes theorem including nuisance parameters takes the form
\begin{equation}
     \mathcal{L}(\theta,\alpha)\times \pi(\theta,\alpha) =\mathcal{P}(\theta,\alpha)\times \mathcal{Z},
    \label{eqn:bayes}
\end{equation}
where we have placed the inputs of inference (likelihood and prior) on the left-hand side, and the outputs (posterior and evidence) on the right. The evidence is as usual equivalent to the (fully-)marginalised likelihood $\mathcal{Z} = \int \mathcal{L}(\theta,\alpha) \pi(\theta,\alpha)d\theta d\alpha$.

We may marginalise any probability distribution so can straightforwardly define the nuisance marginalised posterior and prior by integrating over $\alpha$
\begin{equation}
    \mathcal{P}(\theta) = \int \mathcal{P}(\theta,\alpha)d\alpha, \qquad
    \pi(\theta) = \int \pi(\theta,\alpha)d\alpha.
    \label{eqn:marginal}
\end{equation}
The nuisance marginalised version of Bayes theorem~\cref{eqn:bayes} takes the form
\begin{equation}
    \mathcal{L}(\theta)\times\pi(\theta) = \mathcal{P}(\theta)\times\mathcal{Z}.
    \label{eqn:marginal_bayes}
\end{equation}
Here $\mathcal{Z}$ is the original evidence, whilst $\mathcal{L}(\theta)$ is non-trivially the \emph{nuisance-free likelihood}
\begin{equation}
    \mathcal{L}(\theta) 
\equiv \frac{\int\mathcal{L}(\theta,\alpha)\pi(\theta,\alpha)d\alpha}{\int \pi(\theta,\alpha)d\alpha} = \frac{\mathcal{P}(\theta)\mathcal{Z}}{\pi(\theta)},
    \label{eqn:partial}
\end{equation}
where the above is motivated by marginalising over $\alpha$ of the full Bayes theorem~\cref{eqn:bayes} and substituting the definitions in \cref{eqn:partial,eqn:marginal} recovers the marginalised Bayes theorem~\cref{eqn:marginal_bayes}.

The nuisance-free likelihood \cref{eqn:partial} is straightforward to compute in our framework since we (uniquely) have NS-computed evidences $\mathcal{Z}$ combined with $\textsc{margarine}$ trained distributions $\mathcal{P}(\theta)$ and $\pi(\theta)$ \cite{margarine_neurips} (see \cref{sec:methods}).

We now explain why \cref{eqn:partial} is a useful definition

\begin{Theorem}
    Let $\mathcal{L}_A(\theta,\alpha_A)$ and $\mathcal{L}_B(\theta,\alpha_B)$ be two likelihoods with distinct datasets, each with their own nuisance parameters. The nuisance-free likelihoods $\mathcal{L}_A(\theta)$, $\mathcal{L}_B(\theta)$ form a lossless compression in $\theta$. This means that we can recover the same (marginal) inference in combination that we would have made when performing a combined analysis with all nuisance parameters:
        \begin{align}
            \mathcal{L}_A(\theta,\alpha_A)\mathcal{L}_B(\theta,\alpha_B)\pi_{AB}(\theta,\alpha_A,\alpha_B) &=  \mathcal{P}_{AB}(\theta,\alpha_A,\alpha_B)\mathcal{Z}_{AB},  \label{eqn:combined_bayes}\\
            \Rightarrow
                \mathcal{L}_A(\theta)\mathcal{L}_B(\theta)\pi(\theta) &=  \mathcal{P}_{AB}(\theta)\mathcal{Z}_{AB}, 
                \label{eqn:marginal_combined_bayes}
        \end{align}
 if their respective priors $\pi_A(\theta,\alpha_A)$ and $\pi_B(\theta,\alpha_B)$ satisfy the marginal consistency relations:
        \begin{gather}
            \pi(\theta) = \int \pi_A(\theta,\alpha_A)d\alpha_A = \int \pi_B(\theta,\alpha_B)d\alpha_B,  
            \label{eqn:marginal_A}\\
            \int \pi_{AB}(\theta,\alpha_A,\alpha_B)d\alpha_A = \pi_B(\theta,\alpha_B), \qquad \int \pi_{AB}(\theta,\alpha_A,\alpha_B)d\alpha_B = \pi_A(\theta,\alpha_A).
            \label{eqn:marginal_B}
        \end{gather}                                                                       
        This process is represented graphically in \cref{fig:pipeline} and \cref{fig:pipelineB}.
\end{Theorem}

\begin{proof}
    Integrating the combined Bayes theorem \cref{eqn:combined_bayes} with respect to $\alpha_B$, applying the definition of the marginal posterior \cref{eqn:marginal} on the right-hand side, and drawing out terms independent of $\alpha_B$ on the left,  yields
\begin{equation}
    \mathcal{L}_A(\theta,\alpha_A)\int\mathcal{L}_B(\theta,\alpha_B)\pi_{AB}(\theta,\alpha_A,\alpha_B)d\alpha_B =  \mathcal{P}_{AB}(\theta,\alpha_A)\mathcal{Z}_{AB}.
    \label{eqn:temp1}
\end{equation}
From the definition of a nuisance-free likelihood~\cref{eqn:partial}, and the marginal consistency~\cref{eqn:marginal_B}, we can say that the integral on the left-hand side becomes:
\begin{align}
    &\int \mathcal{L}_B(\theta,\alpha_B) \pi_{AB}(\theta,\alpha_A,\alpha_B)d\alpha_B \nonumber\\
    &= \int \mathcal{L}_B(\theta,\alpha_A,\alpha_B) \pi_{AB}(\theta,\alpha_A,\alpha_B)d\alpha_B 
    &\left[\text{$\mathcal{L}_B(\theta,\alpha_B)\equiv \mathcal{L}_B(\theta,\alpha_A,\alpha_B)$ since $\mathcal{L}_B$ indep of $\alpha_A$}\right]
    \nonumber\\
    &= \mathcal{L}_B(\theta,\alpha_A) {\int \pi_{AB}(\theta,\alpha_A,\alpha_B) d\alpha_B} 
    &\left[\text{Using \cref{eqn:partial} for $\mathcal{L}_B$}\right]
    \nonumber\\
    &= \mathcal{L}_B(\theta) {\int \pi_{AB}(\theta,\alpha_A,\alpha_B) d\alpha_B} 
    &\left[\text{$\mathcal{L}_B(\theta,\alpha_A)\equiv \mathcal{L}_B(\theta)$ since $\mathcal{L}_B$ indep of $\alpha_A$}\right]
    \nonumber\\
    &= \mathcal{L}_B(\theta) \pi_A(\theta,\alpha_A).
    &\left[\text{Using marginal consistency \cref{eqn:marginal_B}}\right]
    \label{eqn:temp2}
\end{align}
Substituting \cref{eqn:temp2} back into \cref{eqn:temp1} we find
\begin{equation}
    \mathcal{L}_A(\theta,\alpha_A)\mathcal{L}_B(\theta)\pi_A(\theta,\alpha_A) =  \mathcal{P}_{AB}(\theta,\alpha_A)\mathcal{Z}_{AB}.
\end{equation}
Proceeding with a similar manipulation to \cref{eqn:temp2}, marginalising with respect to $\alpha_A$, and applying the definition of the nuisance-free likelihood $\mathcal{L}_A(\theta)$ \cref{eqn:partial} and the marginal prior consistency~\cref{eqn:marginal_A} we recover \cref{eqn:marginal_combined_bayes}
\begin{equation}
    \mathcal{L}_A(\theta)\mathcal{L}_B(\theta)\pi(\theta) =  \mathcal{P}_{AB}(\theta)\mathcal{Z}_{AB}. \nonumber
\end{equation}
\end{proof}

\subsection*{Discussion} 
    \Cref{eqn:combined_bayes} represents Bayes theorem for the combined likelihood of both  datasets $\mathcal{L}_{AB}(\theta,\alpha_A,\alpha_B) = \mathcal{L}_A(\theta,\alpha_A)\mathcal{L}_B(\theta,\alpha_B)$, using the combined prior $\pi_{AB}(\theta,\alpha_A,\alpha_B)$. We assume the combined prior is marginally consistent, \cref{eqn:marginal_A,eqn:marginal_B}, which is reasonable, merely demanding that the priors are identical in the parameter spaces where they overlap. In practice, this would usually be achieved by assuming separability between signal and nuisance parameter spaces $\pi(\theta,\alpha) = \pi(\theta)\pi(\alpha)$, but \cref{eqn:marginal_A,eqn:marginal_B} are a slightly less restrictive requirement and therefore more general. 
    
    The upshot of this is that if you have performed inference for two datasets separately, such that you are able to compute the nuisance-free likelihoods with \textsc{margarine}, you may discard the nuisance parameters for the next set of analyses when you combine the datasets.

\section{Methods}
\label{sec:methods}

\textsc{margarine} was first introduced in \cite{margarine_neurips} and uses density estimation to approximate probability distributions such as $\mathcal{P}(\theta)$ and $\pi(\theta)$ given sets of representative samples. The code was developed initially to calculate marginal Kullback-Leibler~(KL) divergences \cite{kullback_information_1951} and Bayesian Model Dimensionalities~(BMD) \cite{Handley_dimensionality_2019} however as discussed in \cref{sec:theory} it can be used to calculate the nuisance-free likelihoods. This in turn means that we can use \textsc{margarine} alongside an implementation of the nested sampling algorithm to sample the product $\mathcal{L}_A(\theta)\mathcal{L}_B(\theta)$. In this manner, \textsc{margarine} allows us to combine constraints on common parameters across different data sets.

We refer the reader to \cite{margarine_neurips} for a complete discussion of how \textsc{margarine} works, however, we discuss briefly the density estimation here. \textsc{margarine} uses two different types of density estimator to model posterior and prior samples, namely Masked Autoregressive Flows~(MAFs, \cite{Papamarkarios_MAF_2017}) and Kernel Density Estimators~(KDEs, \cite{parzen_KDE_1962,rosenblatt_KDE_1956}).

\begin{figure}
\centering
\begin{tikzpicture}[squarednodeA/.style={rectangle, draw=red!60, fill=red!5, very thick, minimum size=5mm},
squarednodeB/.style={rectangle, draw=blue!60, fill=blue!5, very thick, minimum size=5mm},
squarednodeC/.style={rectangle, draw=green!60, fill=green!5, very thick, minimum size=5mm}]

\node[squarednodeA, text width=3cm, align=center](inference3) at (17, -1.5) {Nested Sampling with $\theta$, $\alpha_A$ and $\alpha_B$};

\node[squarednodeB](fulllikelihood1) at (15, 1.5){$ \mathcal{L}_A(\theta,\alpha_A)$};
\node[squarednodeB](fulllikelihood2) at (16.85, 1.5){$ \mathcal{L}_B(\theta,\alpha_B)$};
\node[squarednodeB](fulljointlikelihood) at (16, 0){$ \mathcal{L}_A(\theta,\alpha_A) \mathcal{L}_B(\theta,\alpha_B)$};
\node[squarednodeB](fullprior) at (19, 1.5){$ \pi_{AB}(\theta,\alpha_A, \alpha_B)$};

\draw[->](fulllikelihood1.south) -- (15.5, 0.3);
\draw[->](fulllikelihood2.south) -- (16.5, 0.3);
\draw[->](fullprior.south) -- (inference3.north);
\draw[->](fulljointlikelihood.south) -- (inference3.north);

\node[squarednodeB](jointEvidence2) at (15, -3){$ \mathcal{Z}_{AB}$};
\node[squarednodeB](jointPosterior2) at (17, -3){$ \{\theta\}_{\mathcal{P}_{AB}}$};

\draw[<-](jointEvidence2.north) -- (16, -2);
\draw[<-](jointPosterior2.north) -- (inference3.south);

\node[squarednodeB](jointPosteriorNuisance) at (19, -3){$ \{\alpha_A, \alpha_B\}_{\mathcal{P}_{AB}}$};
\draw[->](18, -2) -- (jointPosteriorNuisance.north);
\draw[blue,thick](16.2, -3.5) -- (20.1, -3.5) -- (20.1, -2.5) -- (16.2, -2.5) -- (16.2, -3.5);

\end{tikzpicture}
\caption{A graphical representation of combining constraints from different data sets via a full nested sampling run over both cosmological and nuisance parameters (\cref{eqn:combined_bayes}).}
\label{fig:pipeline}
\end{figure}
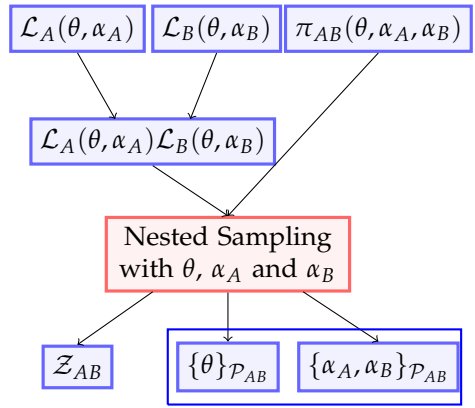

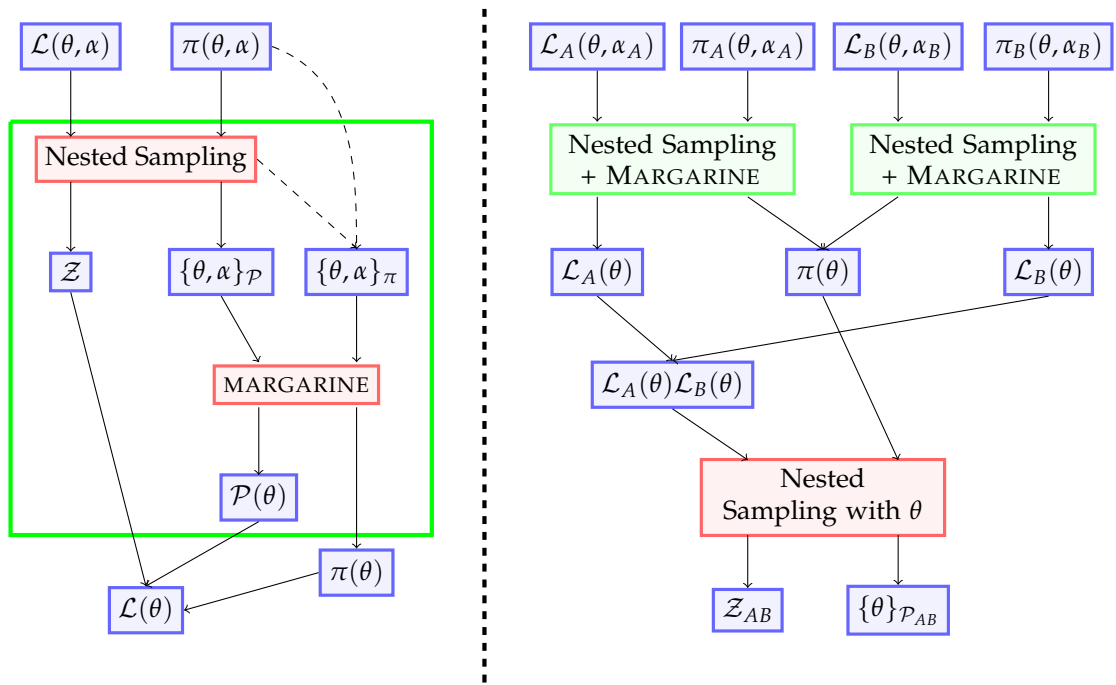
\begin{figure}
\centering
\begin{tikzpicture}[squarednodeA/.style={rectangle, draw=red!60, fill=red!5, very thick, minimum size=5mm},
squarednodeB/.style={rectangle, draw=blue!60, fill=blue!5, very thick, minimum size=5mm},
squarednodeC/.style={rectangle, draw=green!60, fill=green!5, very thick, minimum size=5mm}]
\node[squarednodeA](inference) at (0, 0) {Nested Sampling};
\node[squarednodeB](likelihood) at (-1, 1.5){$ \mathcal{L}(\theta,\alpha)$};
\node[squarednodeB](prior) at (1, 1.5){$ \pi(\theta,\alpha)$};
\node[squarednodeB](evidence) at (-1, -1.5){$ \mathcal{Z}$};
\node[squarednodeB](posterior) at (1, -1.5){$ \{\theta,\alpha\}_\mathcal{P}$};
\node[squarednodeB](priorSamples) at (2.8, -1.5){$ \{\theta,\alpha\}_\pi$};
\node[squarednodeA](margarine) at (2, -3) {\textsc{margarine}};
\node[squarednodeB](marginalPrior) at (2.8, -5.5){$ \pi(\theta)$};
\node[squarednodeB](marginalPosterior) at (1.5, -4.5){$ \mathcal{P}(\theta)$};
\node[squarednodeB](marginalLikelihood) at (0, -6){$ \mathcal{L}(\theta)$};

\draw[green, ultra thick] (-1.8, 0.5) -- (-1.8, -5) -- (3.8, -5)-- (3.8, 0.5) -- (-1.8, 0.5);

\draw[->](likelihood.south) -- (-1, 0.3);
\draw[->](prior.south) -- (1, 0.3);
\draw[->](-1, -0.3) -- (evidence.north);
\draw[->](1, -0.3) -- (posterior.north);
\draw[->](posterior.south) -- (1.5, -2.7);
\draw[dashed,->](prior.east) to[out=-20, in=90] (2.8, -1.2);
\draw[->](priorSamples.south) -- (2.8, -2.7);
\draw[->](2.8, -3.3) -- (marginalPrior.north);
\draw[->](1.5, -3.3) -- (1.5, -4.2);
\draw[->](evidence.south) -- (marginalLikelihood.north);
\draw[->](marginalPosterior.south) -- (marginalLikelihood.north);
\draw[->](marginalPrior.west) -- (marginalLikelihood.east);
\draw[dashed,->](inference.east) -- (priorSamples.north);

\draw[black, ultra thick, dashed] (4.5, 2) -- (4.5, -7);

\node[squarednodeB](likelihood1) at (6, 1.5){$ \mathcal{L}_A(\theta,\alpha_A)$};
\node[squarednodeB](prior1) at (8, 1.5){$ \pi_A(\theta,\alpha_A)$};

\node[squarednodeC, text width=3cm, align=center](NestedMarg1) at (7, 0){Nested Sampling + \textsc{Margarine}};

\draw[->](likelihood1.south) -- (6, 0.5);
\draw[->](prior1.south) -- (8, 0.5);

\node[squarednodeB](marglike1) at (6, -1.5){$ \mathcal{L}_A(\theta)$};
\node[squarednodeB](margprior1) at (9, -1.5){$ \pi(\theta)$};

\draw[->](6, -0.5) -- (6, -1.2);
\draw[->](8, -0.5) -- (9, -1.2);

\node[squarednodeB](likelihood2) at (10, 1.5){$ \mathcal{L}_B(\theta,\alpha_B)$};
\node[squarednodeB](prior2) at (12, 1.5){$ \pi_B(\theta,\alpha_B)$};

\draw[->](likelihood2.south) -- (10, 0.5);
\draw[->](prior2.south) -- (12, 0.5);

\node[squarednodeC, text width=3cm, align=center](NestedMarg1) at (11, 0){Nested Sampling + \textsc{Margarine}};

\node[squarednodeB](marglike2) at (12, -1.5){$ \mathcal{L}_B(\theta)$};
\draw[->](12, -0.5) -- (12, -1.2);
\draw[->](10, -0.5) -- (9, -1.2);

\node[squarednodeB](combinedlike) at (7, -3){$ \mathcal{L}_A(\theta) \mathcal{L}_B(\theta)$};

\draw[->](marglike1.south) -- (combinedlike.north);
\draw[->](marglike2.south) -- (combinedlike.north);

\node[squarednodeA, text width=3cm, align=center](inference2) at (9, -4.5) {Nested Sampling with $\theta$};

\draw[->](combinedlike.south) -- (8, -4);
\draw[->](margprior1.south) -- (10, -4);

\node[squarednodeB](jointEvidence) at (8, -6){$ \mathcal{Z}_{AB}$};
\node[squarednodeB](jointPosterior) at (10, -6){$ \{\theta\}_{\mathcal{P}_{AB}}$};

\draw[->](8, -5) -- (jointEvidence.north);
\draw[->](10, -5) -- (jointPosterior.north);


\end{tikzpicture}
\caption{A graphical representation of combining constraints from two data sets via \textsc{margarine} (\cref{eqn:marginal_combined_bayes}). Left of the dashed line illustrates the derivation of a nuisance-free likelihood function for one experimental data set.}
\label{fig:pipelineB}
\end{figure}

MAFs transform a multivariate base distribution, the standard normal, into a target distribution via a series of shifts and scaling which are estimated by autoregressive neural networks. To improve the performance of the MAF the samples representing the target distribution, in our case $\mathcal{P}(\theta)$ and $\pi(\theta)$, are transformed into a gaussianized space. We implement the MAFs using \textsc{tensorflow} and the \textsc{keras} backend \cite{tensorflow2015-whitepaper}.

KDEs use a kernel to approximate the multivariate probability density of a series of samples. In our case, the kernel is Gaussian and the probability density is a sum of Gaussians centred on the sample points with a given bandwidth. Again we transform the target samples into a gaussianized parameter space allowing the KDE to better capture the distribution. The KDEs are implemented with \textsc{SciPy} in \textsc{margarine} \cite{rosenblatt_KDE_1956, parzen_KDE_1962, 2020SciPy-NMeth}.

Since both types of density estimator build approximations to the target distribution using known distributions, the approximate log-probabilities of the target distribution can be easily calculated. 

The evaluation of normalized log-probabilities for the marginal posterior and marginal prior allows us to calculate the nuisance-free likelihoods, as discussed, along with marginal Kullback-Leibler divergences
\begin{equation}
    \mathcal{D}(\mathcal{P}||\pi) = \int \mathcal{P}(\theta) \log\frac{\mathcal{P}(\theta)}{\pi(\theta)} d\theta,
    \label{eq:kl_divergence}
\end{equation}
which quantifies the amount of information gained when moving from the marginal prior to posterior.

\section{Cosmological Example}
\label{sec:results}

It has previously been demonstrated that \textsc{margarine} is capable of replicating complex probability distributions and approximating marginal Bayesian statistics such as the KL divergence and the BMD \cite{margarine_neurips}. Here we demonstrate the theory discussed in \cref{sec:theory} by combining samples from the Dark Energy Survey~(DES) Year 1 posterior \cite{DES_Year1_2018} and Planck posterior \cite{Planck_cosmo_2020} using \textsc{margarine} to estimate nuisance-free likelihoods. DES surveys supernovae, galaxies and large scale cosmic structures in the universe in an effort to measure dark matter and dark energy densities and model the dark energy equation of state. In contrast, Planck mapped the anisotropies in the Cosmic Microwave Background~(CMB) and correspondingly provided constraints on key cosmological parameters.

\begin{figure}
    \centering
    \includegraphics{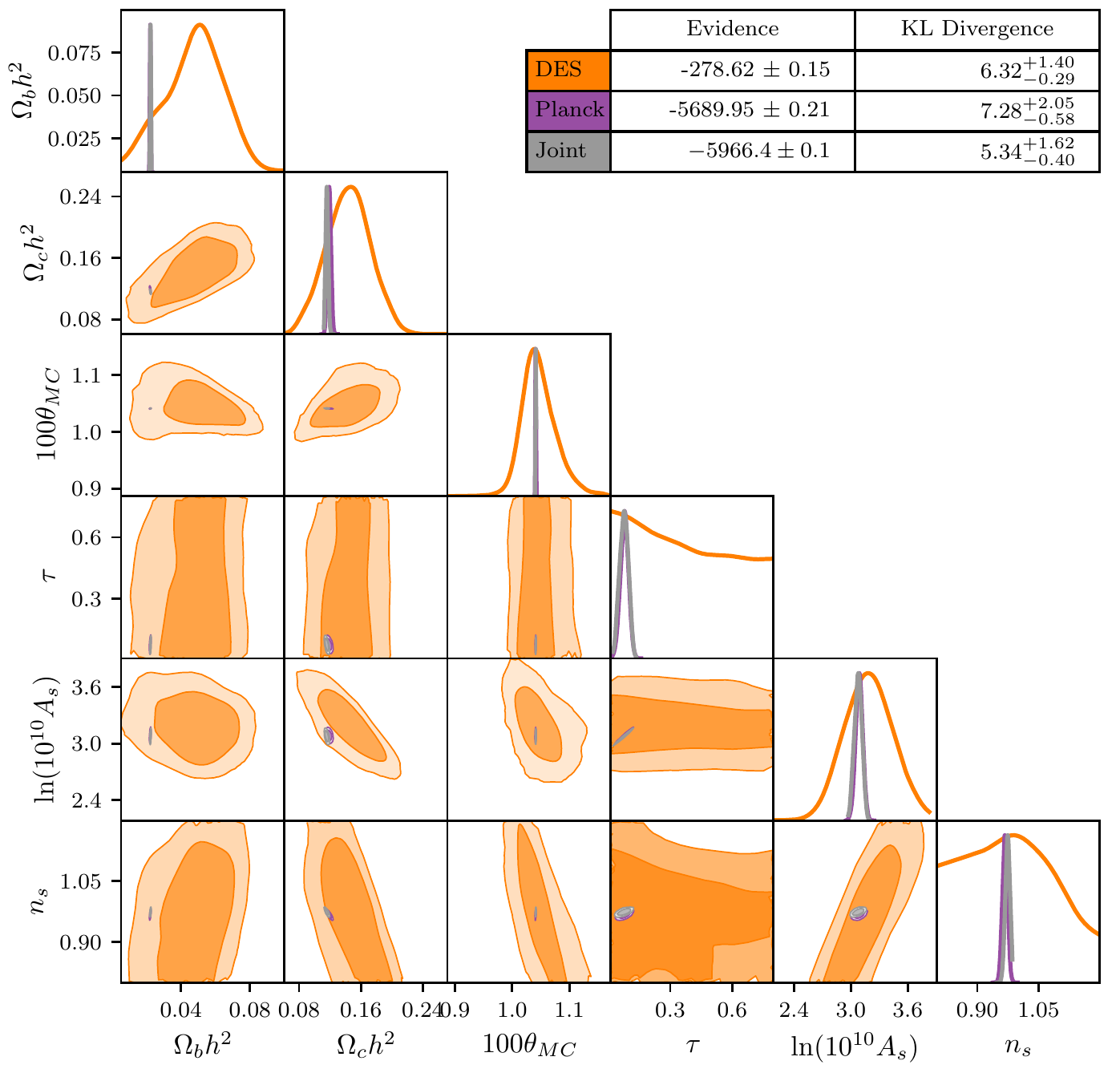}
    \caption{The combined posterior (in grey) found when combining the constraints on the cosmological parameters from DES and Planck using \textsc{margarine}. For DES and Planck, we calculate the marginal KL divergences using \textsc{margarine}, whereas the Bayesian evidences are calculated using \textsc{anesthetic}. The joint evidence and joint KL divergence are calculated with a combination of the two codes and are found to be approximately consistent with those found in the literature \cite{Handley_tensions_2019, Handley_dimensionality_2019}. Note that the error on the joint evidence is likely underestimated as it relies on evaluations of log-probabilities for various distributions, for which \textsc{margarine} does not currently provide errors. The figure produced with \textsc{anesthetic} \cite{anesthetic}.}
    \label{fig:joint}
\end{figure}

The constraints from DES and Planck have previously been combined using a full nested sampling run over all parameters including a multitude of `nuisance' parameters in a computationally expensive exercise \cite{Handley_tensions_2019}. This corresponds to the flow chart in \cref{fig:pipeline} and the previous analysis gives us access to the combined DES and Planck evidence, which is found to have a value of $\log(\mathcal{Z}) = -5965.7 \pm 0.3$. In \cref{fig:joint} we show the DES, Planck and joint posteriors for the six cosmological parameters derived in this work using \textsc{margarine} and the flow chart in \cref{fig:pipelineB}. The constrained parameters are the baryon and dark matter density parameters, $\Omega_b h^2$ and $\Omega_c h^2$, the angular size of the sound horizon at recombination, $\theta_{MC}$, the CMB optical depth, $\tau$, the amplitude of the power spectrum, $A_s$, and the corresponding spectral index, $n_s$. These make up the set $\theta = (\Omega_b h^2, \Omega_c h^2, \theta_{MC}, \tau, A_s, n_s)$. We use the nested sampling algorithm \textsc{polychord} in our analysis \cite{Handley2015a, Handley2015b}.

We use a uniform prior that is defined to be three sigma around the Planck posterior mean. This is done to improve the efficiency of our nested sampling run. However, we subsequently have to re-weight the samples and correct the evidence for the difference between the priors used here and in the previous full nested sampling run \cite{Handley_tensions_2019} for comparison. If we define
\begin{equation}
    \mathcal{Z}_A = \int \mathcal{L}(\theta) \pi_A(\theta) d\theta, \qquad
    \mathcal{Z}_B = \int \mathcal{L}(\theta) \pi_B(\theta) d\theta, 
\end{equation}
where $A$ is our uniform prior space and $B$ is our target prior space from the previous work \cite{Handley_tensions_2019}, then
\begin{equation}
    \mathcal{Z}_B 
    = \int \mathcal{L}(\theta) \pi_B(\theta) d\theta
    = \int \mathcal{L}(\theta) {\pi_A(\theta)}\frac{\pi_B(\theta)}{\pi_A(\theta)} d\theta 
   = \mathcal{Z}_A\left\langle \frac{\pi_B(\theta)}{\pi_A(\theta)}\right\rangle_{\mathcal{P}_A}
\end{equation}
giving
\begin{equation}
    \mathcal{Z}_B 
    = \mathcal{Z}_A\left\langle \frac{\pi_B(\theta)}{\pi_A(\theta)}\right\rangle_{\mathcal{P}_A}.
\end{equation}
Then following similar arguments we can transform our posteriors by re-weighting the distributions with the following
\begin{equation}
    w^{(i)}_B 
    = w^{(i)}_A  \frac{\pi_B(\theta^{(i)})}{\pi_A(\theta^{(i)})}.
\end{equation}

We see from the figure and corresponding table that with our joint analysis we are able to derive a log-evidence that is approximately consistent with that found in \cite{Handley_tensions_2019} validating the theory discussed and its implementation with \textsc{margarine}. We note that the re-weighting described above relies on calculation of the two prior log-probabilities for which we use \textsc{margarine} and currently do not have an estimate of the error for. As a result, the error in the combined evidence, $Z_B$, is given by the error in $Z_A$ from the nested samples and is likely underestimated. Using \textsc{margarine} \cite{margarine_neurips} we can also derive the combined KL divergence, also reported in \cref{fig:joint}, which we find is consistent with the result in the literature of $\mathcal{D} = 6.17 \pm 0.36$ \cite{Handley_dimensionality_2019}. Similarly, we derive marginal KL divergences for the DES and Planck cosmological parameters using \textsc{margarine}. A full discussion of the implications of combining the two data sets for our understanding of cosmology can be found in the literature \cite[e.g][]{Handley_tensions_2019, Handley_dimensionality_2019}.

By reducing the number of parameters that need to be sampled, we significantly reduce the nested sampling runtime. For \textsc{polychord} the runtime scales as the cube of the number of dimensions \cite{supernest}. This can be seen by assessing the time complexity of the algorithm where, $T \propto n_\mathrm{live} \times \langle T\{\mathcal{L}(\theta)\}\rangle \times \langle T\{Impl.\}\rangle \times \mathcal{D}(\mathcal{P}||\pi)$. Here $n_\mathrm{live}$ scales with the number of dimensions, $d$, as does the Kullback-Leibler divergence. For \textsc{polychord}, the specific implementation time complexity factor, $\langle T\{Impl.\}\rangle$, representing the impact of replacing dead points with higher likelihood live points on the runtime, scales linearly with $d$. Together this gives $T \propto d^3 \times \langle T\{\mathcal{L}(\theta)\}\rangle$. Therefore, by using nuisance-free likelihoods and sampling over 6 parameters rather than 41 parameters (cosmological plus 20 nuisance parameters for DES and 15 different nuisance parameters for Planck) we reduce the runtime by a factor of $(41/6)^3 \approx 319$ with further improvements in $\langle T\{\mathcal{L}(\theta)\}\rangle$. Using \textsc{margarine}, $\langle T\{\mathcal{L}(\theta)\}\rangle$ is typically reduced since analytic likelihoods are computationally more expensive than emulated likelihoods.

\section{Conclusion}
\label{sec:conclusion}

In the paper, we have demonstrated the consistency between combining constraints from different experiments in a marginal framework using density estimators and the code \textsc{margarine} with a full nested sampling run over all parameters, including those describing `nuisance' components of the data. We have shown this consistency mathematically and with a cosmological example. For the combination of Planck and DES, we find a Bayesian evidence and KL divergence that is consistent with previous results \cite{Handley_tensions_2019, Handley_dimensionality_2019}.

The analysis in this paper is only possible because (a) we are able to estimate densities in the (much smaller) cosmological parameter space $\theta$ using \textsc{margarine}, and (b) because we have evidences, $\mathcal{Z}$, from our original nested sampling runs. It is this unique combination which allows us to compress away or discard nuisance parameters once they have been used. We note also that working in the marginal space results in a compression that is lossless in information on $\theta$ as it recovers an identical marginal posterior and total evidence as is found during a full Bayesian inference. Finally, through the nuisance-free likelihood we can significantly reduce the dimensionality of our problems and since it is faster to emulate a likelihood rather than analytically evaluate, \textsc{margarine} offers a much more computationally efficient path to combined Bayesian analysis.

In principle, our work paves the way for the development of a publicly available library of cosmological density estimators modelled with \textsc{margarine} that can be combined with new data sets using the proposed method in a more efficient manner than currently implemented techniques. However, the work has implications outside of cosmology in any field where multiple experiments probe different aspects of the same physics.





\authorcontributions{Conceptualization, W.H. and H.B.; methodology, W.H. and H.B.; formal analysis, H.B.; software, H.B.; writing---original draft preparation, H.B.; writing---review and editing, H.B., W.H., P.L. and P.S.; supervision, W.H., E.d.L.A. and A.F.. All authors have read and agreed to the published version of the manuscript.}

\funding{H.B. acknowledges the support of the Science and Technology Facilities Council (STFC) through grant number ST/T505997/1 and Fitzwilliam College, Cambridge. W.H. and A.F. were supported by Royal Society University Research Fellowships. PHS acknowledges support from a McGill Space Institute Fellowship and the Canada 150 Research Chairs Program. E.d.L.A. was supported by the STFC through the Ernest Rutherford Fellowship.}


\conflictsofinterest{The authors declare no conflict of interest.}

\reftitle{References}

\bibliography{refs, journals}

\end{document}